# Kicking the Can Down the Road: Understanding the Effects of Delaying the Deployment of Stratospheric Aerosol Injection

Ezra Brody[1], Daniele Visioni[2], Ewa M. Bednarz[1,3,4], Ben Kravitz[5,6], Douglas G. MacMartin[1], Jadwiga H. Richter[7], Mari R. Tye[7,8]

[1]Department of Mechanical and Aerospace Engineering, Cornell University, Ithaca, NY, USA
[2]Department of Earth and Atmospheric Sciences, Cornell University, Ithaca, NY, USA
[3]Cooperative Institute for Research in Environmental Sciences (CIRES), University of Colorado Boulder, Boulder, CO, USA
[4]NOAA Chemical Sciences Laboratory, Boulder, CO, USA
[5]Department of Earth and Atmospheric Sciences, Indiana University, Bloomington, IN, USA
[6]Atmospheric Sciences and Global Change Division, Pacific Northwest National Laboratory, Richland, WA, USA
[7]National Center for Atmospheric Research, Boulder, CO, USA
[8]Whiting School of Civil Engineering, Johns Hopkins University, Baltimore, MD, USA

**Correspondence:** Ezra Brody (eb637@cornell.edu)

## Abstract

Climate change is a prevalent threat, and it is unlikely that current mitigation efforts will be enough to avoid unwanted impacts. One potential option to reduce climate change impacts is the use of stratospheric aerosol injection (SAI). Even if SAI is ultimately deployed, it might be initiated only after some temperature target is exceeded. The consequences of such a delay are assessed herein. This study compares two cases, with the same target global mean temperature of ~1.5°C above preindustrial, but start dates of 2035 or a "delayed" start in 2045. We make use of simulations in the Community Earth System Model version 2 with the Whole Atmosphere Coupled Chemistry Model version 6 (CESM2-WACCM6), using SAI under the SSP2-4.5 emissions pathway. We find that delaying the start of deployment (relative to the target temperature) necessitates lower net radiative forcing (-30%) and thus larger sulfur dioxide injection rates (+20%), even after surface temperatures converge, to compensate for the extra energy absorbed by the Earth system. However, many of the surface climate differences between the 2035 and 2045 start simulations appear to be small during the 10-25 years following the delayed SAI start, although longer simulations would be needed to assess any longer-term impacts in this model. In addition, irreversibilities and tipping points that might be triggered during the period of increased warming may not be adequately represented in the model but could change this conclusion in the real world.

I. Introduction

Increased concentrations of carbon dioxide and other greenhouse gasses (GHG) have been causing global mean temperatures to rise. If left uncurtailed, this could lead to catastrophic outcomes to ecosystems and human lives throughout the world. Efforts are being made to decrease carbon dioxide emissions, but there are several technological and societal barriers that limit progress in this regard. These barriers make it unlikely that the world will stay beneath the 1.5-degree-above-preindustrial temperature target set by the Paris Agreement (IPCC 2018). Stratospheric aerosol injection (SAI) has been gaining attention as a potential method to avoid catastrophic effects of climate change while efforts are made to decrease greenhouse gas concentrations. SAI would involve lofting aerosols - typically sulfate - or their precursors into the lower stratosphere. This would reflect a small amount of incoming solar radiation back to space, cooling the planet as a result.

There is a vast array of potential scenarios in which SAI could be deployed, making analysis of possible outcomes difficult (MacMartin et al. 2022). These variables include, but are not limited to, the underlying GHG emission scenario (e.g. Tilmes et al., 2020), the start date of a potential SAI deployment, and a desired level of cooling or temperature target. Even with just the latter two variables, there are many permutations that could occur. For a fixed start date, one could choose different levels of desired cooling - or temperature targets - and analyze the change in outcomes. This has been studied by Visioni et al. 2023 using the middle-of-the-road SSP2-4.5 GHG emission scenario and SAI starting in 2035. On the other hand, one could choose a single SAI temperature target and vary the start date of SAI deployment. The latter is the focus of this paper, where we consider the differences in outcomes between starting SAI deployment roughly when surface temperatures reach the target, versus delaying and using SAI to cool back down to target temperature. A similar question has recently been studied by Pflüger et al. 2023, who analyzed an extreme example of a sixty-year delay under a high-emissions SSP5-8.5 scenario to maximize the signal-to-noise ratio; they focused primarily on the lag of ocean dynamics and inability to recover the Atlantic Meridional Overturning Current (AMOC). The goal of this study is to analyze the broader impacts of a more moderate delayed-start scenario - starting SAI deployment in 2045 vs 2035 under SSP2-4.5 emissions. These simulations have also been used by Hueholt et al. 2023 to study the ecological impacts of the rapid cooling associated with the delayed start case, whereas we focus more on the differences between the 2035 and 2045 start-date cases. We use the original ARISE dataset described in Richter et al 2022, as well as two new datasets; these extended-ARISE datasets are available to the broader research community for analysis.

Understanding the effects of waiting longer to start deployment is important for policy-makers. Stratospheric aerosol injection research is a young field, and there are still many important knowledge gaps (e.g., NASEM 2021) that scientists are working to address. Additionally, it may take considerable time to develop the capability to deploy SAI as it would require novel aircraft (Smith & Wagner 2018). Perhaps most importantly, the institutional capacity to make a decision on deployment as well as oversee its implementation does not yet exist. However, the downsides of waiting longer might be significant, so it is important to factor this into a decision on whether or not to deploy, or start developing the capability to deploy, at any given time.

There are two broad and related considerations associated with a delayed deployment to restore to an earlier temperature: whether the same climate state is eventually achieved as an earlier deployment scenario, and the climate damages and risks incurred during the time interval before an SAI deployment achieves the desired target temperature. One reason why the same climate state may not be achieved in a delayed start case is the existence of irreversible elements in the climate system, commonly referred to as tipping points. These are changes to the state of the climate system that, once triggered, cannot be reversed by simply reversing the forcing mechanism (McKay et al. 2022). Such events include shifting to a different regime of the Atlantic Meridional Overturning Circulation (AMOC) (Ditlevsen & Ditlevsen 2023), warming-driven species extinction, ice sheet mass loss, or a reduction in the productivity in the Amazon forest (Amazon Dieback). Knowing whether delaying the start of SAI deployment could result in one of these tipping elements being triggered would be an important factor to consider when making a decision about whether to deploy at any given time.

One potential benefit of delaying SAI deployment is the recovery of ozone due to the decrease in halogens in the stratosphere. SAI exacerbates the chemical ozone depleting effects of halogens as the aerosols provide active surfaces on which heterogeneous reactions can convert halogens from inactive reservoir forms (e.g. HCl, $ClONO_2$) into active, ozone-destroying forms (e.g. $Cl_2$, ClO) (Haywood & Tilmes 2022). Stratospheric halogen concentrations have been decreasing since the 1990s and are projected to continue to decrease for the next century as a result of the phase-out of the long-term ozone depleting substances introduced by the 1987 Montreal Protocol and its subsequent amendments and adjustments. It follows that waiting longer to start deploying SAI could minimize the negative impact on the ozone layer.

Delayed deployment will also lead to higher injection rates being needed even for the same ultimate temperature target, as sufficient aerosols would be needed to manage the increased energy accumulated in the Earth system during the delay. This increases cost. But more importantly, this would exacerbate the physical and chemical effects related to increased sulfate load in the atmosphere. These could include stratospheric heating and the resulting modulation of stratospheric circulation (e.g. Simpson et al. 2019; Bednarz et al. 2023a), weakening of the Hadley cell (Cheng et al. 2022), modulation of the extratropical modes of variability like Northern and Southern Annular Modes (Banerjee et al. 2020; Banerjee et al. 2021; Bednarz et al. 2022 ), ozone depletion (Tilmes et al. 2022), changes in the hydrological cycle (Simpson et al. 2019; Tye et al. 2022; Tew et al. 2023), and acid rain impacts (Visioni et al. 2020). Furthermore, the relationship between injection rate and cooling is known to be nonlinear (Visioni et. al. 2023), with diminishing returns for increased injection rates. This will further increase the needed injection rate from a delayed start. It is also important to consider not just whether the climate can eventually be returned to the same state, but the amount of time and the $SO_2$ injection rates that are needed to do so.

## II. Methodology
### A. Model Description

All simulations were run using the Community Earth System Model, version 2 (CESM2), using the Whole Atmosphere Community Climate Model, version 6 (WACCM6) for the atmosphere model. The configuration has a horizontal resolution of 1.25° longitude by 0.9° latitude. It has 70 vertical layers, with a model top at $4.5 \times 10^{-6}$ hPa (~140 km). This high top is important for resolving the stratospheric dynamics that transport the aerosols in SAI. WACCM6 also includes interactive tropospheric-stratospheric-mesospheric-lower-thermospheric chemistry (TSMLT), prognostic aerosols using the Modal Aerosol Model version 4 (MAM4), as well as an interactive sulfur cycle. This allows for representation of $SO_2$ oxidation and subsequent nucleation and coagulation into sulfate aerosols, as well as aerosol transport. These mechanisms are a core component of SAI, so accurate representation of these processes is important for these simulations. A more detailed description of CESM2(WACCM6) can be found in Danabasoglu et al. (2020).

### B. Reference simulations

The reference simulation used in this study is the Shared Socioeconomic Pathway scenario SSP2-4.5, which is the middle-of-the-road emissions scenario that is roughly consistent with current global climate policy (Parson et al. 2007, 2008). A 5-member ensemble of this scenario was simulated with CESM2(WACCM6) as part of the CMIP6 project for the years 2015-2100. An additional shorter 5-member ensemble, covering years 2015-2070, was produced as part of the Assessing Responses and Impacts of Solar climate intervention on the Earth system with Stratospheric Aerosols (ARISE-SAI) simulations (Richter et al. 2022). ARISE-SAI sets out a simulation protocol, following the guidance scenario outlined in MacMartin et al. (2022), that can be replicated by other simulation centers (e.g. Henry et al. 2023). ARISE-SAI-1.5 was the first set of simulations, and designed to maintain a global mean temperature ~1.5°C above preindustrial levels. The simulation protocol recommends a minimum simulation ensemble size of three members, an initial spin-up period of 10 years following deployment and 20 further years of quasi-equilibrium simulation. The nomenclature indicates the nature of the climate intervention, for instance SAI, and the global mean temperature target. In addition to ARISE-SAI-1.5, two further ensembles using CESM2(WACCM) are available for community analysis: ~1.5°C delayed start (ARISE-SAI-DELAY) and ~1.0°C (ARISE-SAI-1.0), described below.

### C. Climate intervention simulations

There are three SAI scenarios examined in this study with CESM2(WACCM6). In all three scenarios sulfate aerosol precursors, namely sulfur dioxide ($SO_2$) gas, is injected into the lower stratosphere with the aim of reducing surface temperatures. For all three scenarios, a 10-member ensemble of simulations was carried out and the simulations use the same non-SAI forcings from the SSP2-4.5 scenario. The differences in the SAI scenarios are in the temperature targets, as explained below, and the start year of SAI. These three scenarios are as follows:(i) a global average temperature target of 1.5°C above preindustrial with SAI starting in 2035 (already described in Richter et al. 2022), (ii) a global average temperature target of 1.0°C above preindustrial with SAI starting in 2035, and (iii) a global average temperature target of 1.37°C above preindustrial with SAI starting in 2045. This third ensemble was originally intended to have been the same temperature target of 1.5°C as the first scenario but was unfortunately run at a slightly lower target; below we linearly interpolate between the first two sets of simulation output to generate a 2035-start dataset that has the same temperature target as this 2045-start case. For the purposes of this study, 1.5°C above preindustrial is defined as the average global-mean near-surface air temperature from 2020-2039 in the first ensemble member of the SSP2-4.5 simulation in CESM2(WACCM6) (see detailed explanation in Richter et al. 2022); 2020-2039 was chosen as representative of the time period when the Earth's global mean temperature is likely to reach 1.5°C above preindustrial (MacMartin et al. 2022). The 1.0°C target corresponds to the 2000-2019 average, and the 1.37°C target in the 2045-start run corresponds to the 2015-2035 average.

All of the simulations use the same SAI injection strategy developed in Kravitz et al 2017 (and used in Tilmes et al 2018, 2020, MacMartin et al 2022, and Richter et al 2022) that employs a feedback algorithm to adjust $SO_2$ injection rates at four different latitudes to maintain three large-scale near-surface air temperature metrics. The four $SO_2$ injection locations used here are 30°S, 15°S, 15°N, and 30°N, all at 0°E, and ~21.5 km altitude, and the three climate outcomes are the global mean temperature ($T_0$), the interhemispheric temperature gradient ($T_1$), and the equator-to-pole temperature gradient ($T_2$). Complete formulas and definitions for $T_1$ and $T_2$, as well as the complete control law used in these simulations, are defined in Kravitz et al. (2017). Values for $T_0$, $T_1$, and $T_2$ used in the controllers for the three scenarios are listed in Table 1; the target values for T1 and T2 are chosen as the average values over the reference period in the SSP2-4.5 simulation consistent with the choice of T0 target. In the 2045 simulation, the initial temperature is significantly above the ultimate target, and a smooth transition was obtained by ramping down the target gradually to those listed in the table over the first five years of the simulation. In the 1.0°C 2035 simulation, the target values were ramped down over the first ten years of the simulation. Injections occur at each timestep in the model; at the end of the year, the controller calculates the new injection rates for the following year.

|  | $T_0$ | $T_1$ | $T_2$ |
| --- | --- | --- | --- |
| **1.5°, 2035** | 288.64 K | 0.8767 K | -0.589 K |
| **1.37°, 2045** | 288.51 K | 0.880 K | -0.587 K |
| **1.0°, 2035** | 288.14 K | 0.849 K | -0.590 K |

*Table 1: Controller targets for the three scenarios. The first 10 years of the 1.37° 2045 and 1.0° 2035 scenarios gradually ramp down to these targets.*

### D. Interpolated simulation

The goal of this study is to isolate the effect of delaying the start of SAI deployment. In order to make a direct comparison between start dates, we need an ensemble of 2035-start but 1.37°C target simulations. We obtain these simulations via interpolation: the data for this scenario is calculated by taking a linear combination of the 2035 start 1.5°C run (74%) and the 2035 start 1.0°C run (26%). While we acknowledge the limitations of this method, as the climate is a nonlinear system, we note that the responses of certain climate variables to a forcing like GHG or SAI have been shown to be reasonably linear with small changes in surface temperature (MacMartin & Kravitz, 2016; IPCC, 2021; Visioni et al., 2023) and as such the method constitutes a reasonable compromise in lieu of having an ensemble of 2035-start 1.37°C target simulations.

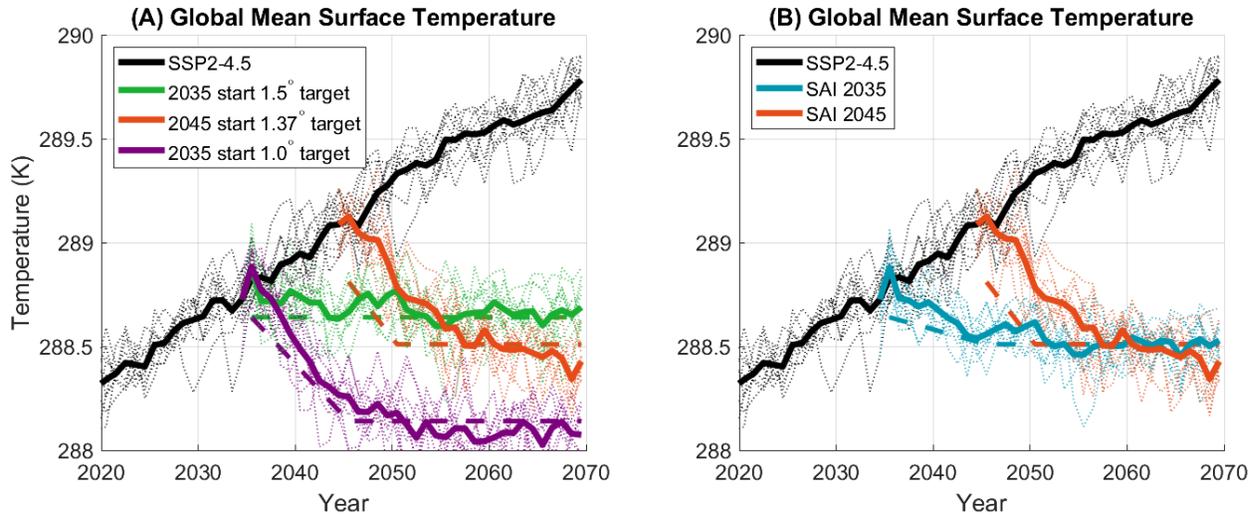

*Figure 1: Global mean surface air temperature vs time. Thick solid lines represent ensemble averages. Thin dotted lines represent individual ensemble members. Thick dashed lines represent the controller targets. (A) Direct Results from the simulations. (B) The scenarios used for this study. "SAI 2035" is a linear interpolation of "2035 start 1.5° target" and "2035 start 1.0° target" in panel A.*

From here on out, the reference simulations with no SAI will be called "SSP2-4.5," the interpolated simulations with a 2035 start and 1.37° target will be called "SAI 2035," and the simulations with a 2045 start and 1.37° target will be called "SAI 2045." Figure 1(B) shows the time evolution of the global mean near-surface air temperature in these simulations. For all subsequent discussions, the reference period refers to the 1.37°-above-preindustrial target, which corresponds to the average over 2015-2035.

### III. Results and Discussion

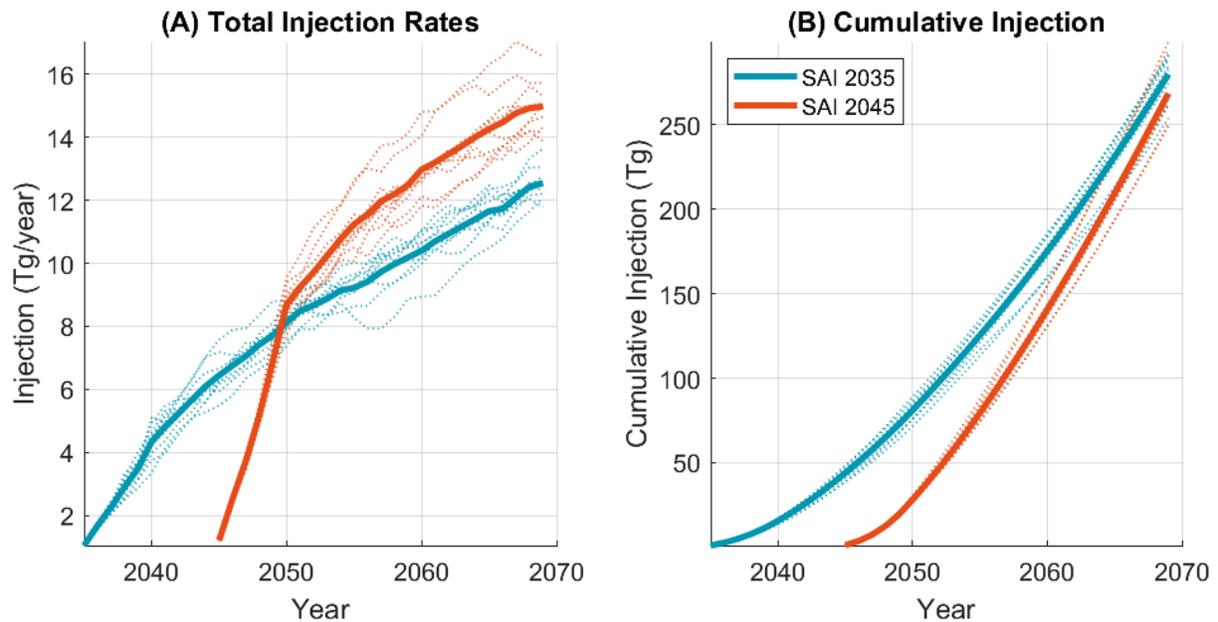

*Figure 2: Left (2A) - Total SO₂ injection rates each year, summed over the four injection latitudes. Right (2B) - cumulative SO₂ injections, summed over the four injection latitudes.*

The timeseries of $SO_2$ injection rates in the SAI 2035 and SAI 2045 scenarios are shown in Figure 2, summed over the 4 injection latitudes. SAI 2045 requires significantly higher injection rates than the SAI 2035 scenario – approximately 20% higher in the last decade of simulation. This is the case even after the initial cooling period (Figure 2A), indicating that even though the global mean surface air temperatures are the same, the overall climate state is not, which is consistent with Pflüger et al (2024). The change in radiative forcing resulting from a change in injection rate is known to be nonlinear due to increased aerosol coagulation at higher injection rates (Visioni et al. 2023), so the exact magnitude of the difference shown with the linear interpolation SAI 2035 scenario may be inaccurate. However, a nonlinear calculation yielded very similar results (see supplemental material). Interestingly, cumulative injections for the two scenarios are similar at the end of the simulations around 2070 (Figure 2B), but given the known nonlinearities and limited simulation duration, it is unclear whether they would converge past 2070. This has implications for the total long-term cost of an SAI program. In addition, the increased stratospheric aerosol loads under higher injection rates in SAI 2045 will lead to larger aerosol-induced stratospheric heating, with implications for stratospheric and tropospheric circulations and, thus, regional surface climate (e.g. Simpson et al. 2019; Bednarz et al. 2023-strategy; Bednarz et al., 2023-nonlinearities), as well as pose increased risks from acid rain as higher stratospheric aerosol loads are removed from the atmosphere (e.g. Visioni et al., 2020).

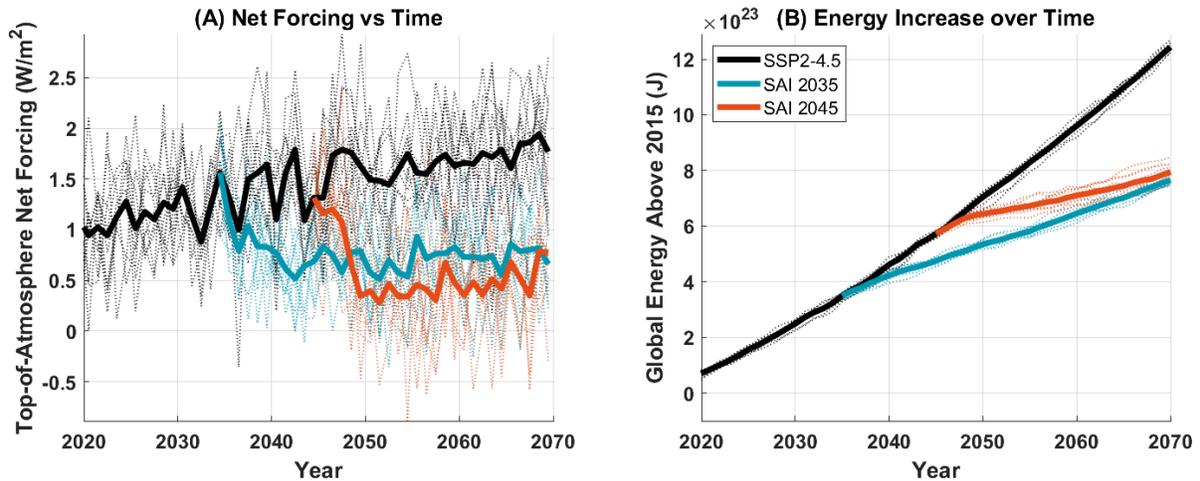

*Figure 3: Left panel (3A) - Total top-of-atmosphere net forcing vs time. Right panel (3B) - Global energy increase above 2015 levels, calculated by integrating the left panel plot over time.*

In both SAI scenarios, the ensemble-average net downward radiative forcing at the top of the atmosphere is positive every year. This means that despite the global-mean surface temperature staying constant towards the end of the SAI simulations, the energy within the climate system is increasing. That is, the Earth is not in thermal equilibrium, with the deep ocean continuing to warm. Furthermore, Figure (3A) shows that after the initial transient, SAI 2045 has an approximately 30% lower net forcing than SAI 2035 due to higher sulfate aerosol loading, despite the same surface temperature. By 2070, the total energy of the climate system in the 2045 scenario is nearly equal to that of the 2035 scenario with the same temperature target.

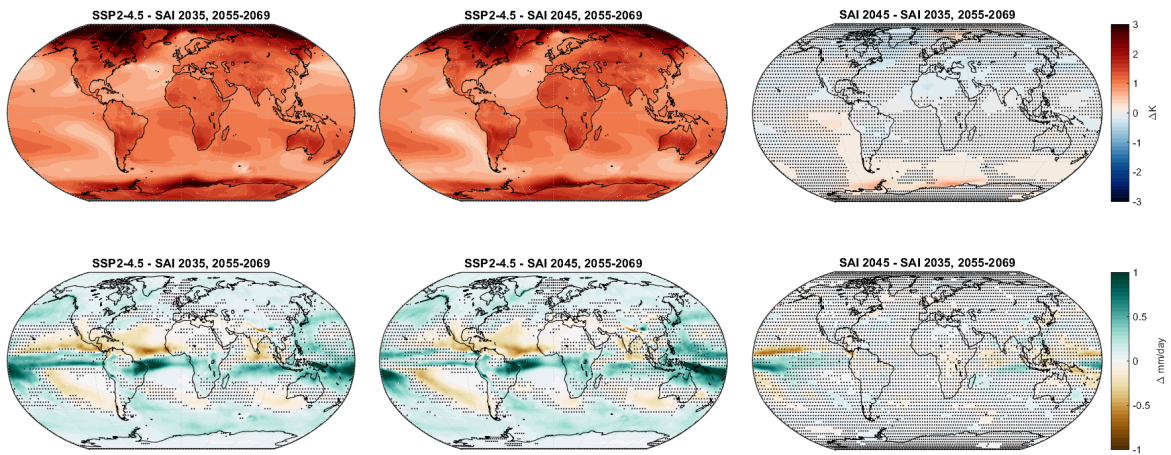

*Figure 4: Average temperature difference from 2055-2069 between: SSP2-4.5 and SAI 2035 (top left), SSP2-4.5 and SAI 2045 (top middle), and SAI 2045 and SAI 2035 (top right). Bottom row is the same but for precipitation. Statistically insignificant areas are stippled.*

Figure 4 shows the spatial differences in surface temperature and precipitation between the scenarios from 2055-2069, after global mean surface temperatures have converged. There are detectable differences in surface temperature in certain regions between SAI 2045 and SAI 2035 in annual mean. These differences are small, however, compared to the differences between either SAI scenario and SSP2-4.5. Figure S8 shows the December-January-February (DJF); there is a slightly larger winter-warming signal over Northern Eurasia under the delayed-start case. This is likely the result of the higher injection rates and the associated stronger tropical stratospheric heating leading to a stronger high latitude response and the associated positive NAO response at the surface (Bednarz et al., 2023a,b.)

Pflüger et al (2024) found a significant difference in the pattern of surface air temperature between their gradual-start and 70-year-delayed case that was a result of an inability of their delayed-case to restart deep-convection and recover the strength of the Atlantic Meridional Overturning Circulation (AMOC). Here we find no statistically significant difference in AMOC strength with our 10-year delay by the end of the simulation (Figure S5), suggesting that the overall length of delay and warming during that period can affect conclusions.

For precipitation, there are very few areas that are statistically significant between SAI 2035 and SAI 2045. They are all small in area, small in magnitude, or over the ocean. Both SAI scenarios show significant differences in precipitation from SSP2-4.5. The main differences are a shift in the tropical precipitation belt and a small but consistent difference in precipitation at high latitudes.

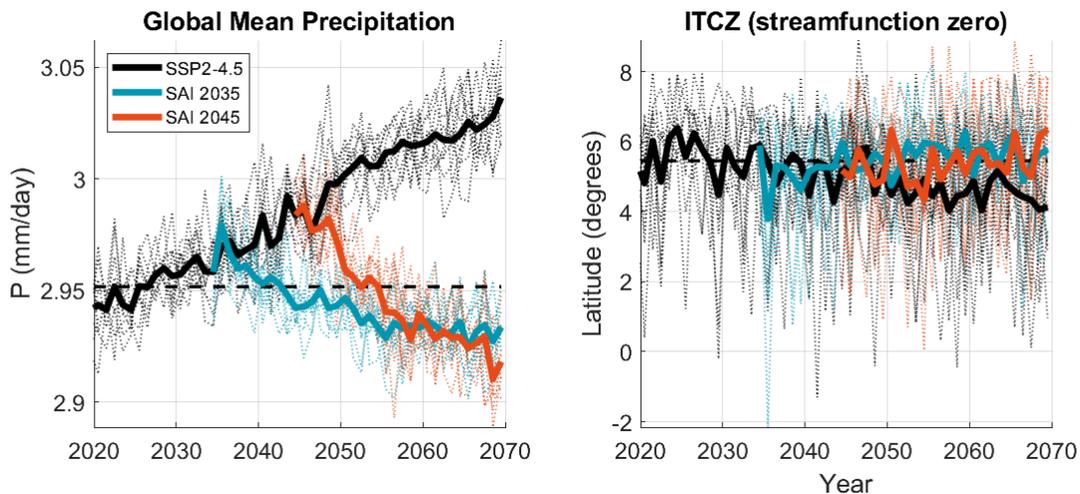

*Figure 5: Global mean precipitation (left) ITCZ as the meridional streamfunction zero (Haigh et al. 2005) (right) vs time for the three scenarios. Black dashed lines are values from the reference period (2015-2035)*

These findings are corroborated by Figure 5. Global mean precipitation increases over time in SSP2-4.5, but decreases to below the reference period level in both SAI scenarios. The two SAI scenarios have very similar amounts of precipitation over the last ~15 years. Likewise, the ITCZ shifts southward in SSP2-4.5, but the location remains relatively steady in both SAI scenarios.

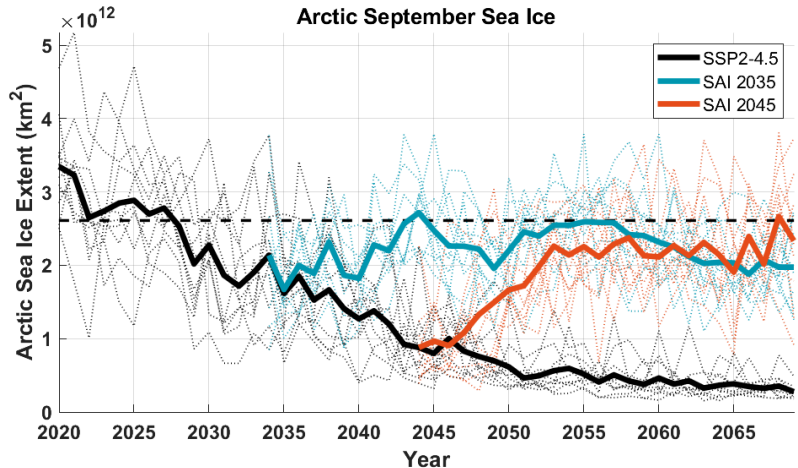

*Figure 6: Plot of Arctic September sea ice vs time for the four simulations. The dashed black line represents the average value from the reference period (2015-2035).*

Figure 6 shows the time evolution of the Arctic September sea ice (SSI) extent for the three scenarios. In general, March sea ice extent over the Arctic is not projected to change significantly over the next several decades under SSP2-4.5 in this model, and the addition of SAI does not change that significantly in any of the simulations (not shown). However, September Arctic sea ice is projected to decrease significantly over the next several decades under the SSP-2.45 scenario, and the CESM2(WACCM6) simulations suggests that SAI can reverse this decline under both SAI scenarios, indicating that at least in this model, the additional sea ice decline between 2035 and 2045 is fully reversible by a delayed-SAI. The difference in SSI extent between the 2035 and 2045 cases from 2055-2069 is not statistically significant at the 95% level (p=.9109). However, in both SAI scenarios SSI extent does not quite reach the same level as in the reference period, and this difference is statistically significant (p values of .0488 and .0251 for the 2035 start and 2045 start cases, respectively, using a t-test with ensemble members as samples).

Figure 7 shows how land carbon north of 50°N changes in the three scenarios. Total ecosystem carbon increases throughout the simulations for all scenarios as increased $CO_2$ and temperatures increase plant life (Lee et al., 2023). Soil carbon increases more over time in SAI scenarios, whereas vegetation carbon increases more in SSP2-4.5. In SAI 2045, this increase in soil carbon lags behind that of SAI 2035. The plots on the bottom of Figure 7 suggest that this difference is mainly a difference in permafrost. Permafrost area decreases throughout SSP2-4.5, but the rate of loss is slowed in SAI 2035. In SAI 2045, permafrost area recovers in the first 15 years or so and reaches similar levels to that of SAI 2035 by 2070. However, the soil carbon lost due to permafrost thaw cannot be recovered by simply refreezing the soil. Figure 7 (bottom right) shows that significant area at high latitudes has less carbon long-term in SAI 2045 than in SAI 2035, indicative of at least 450 Tg of carbon released by permafrost thaw that is not recoverable on these time scales with the later SAI start date.

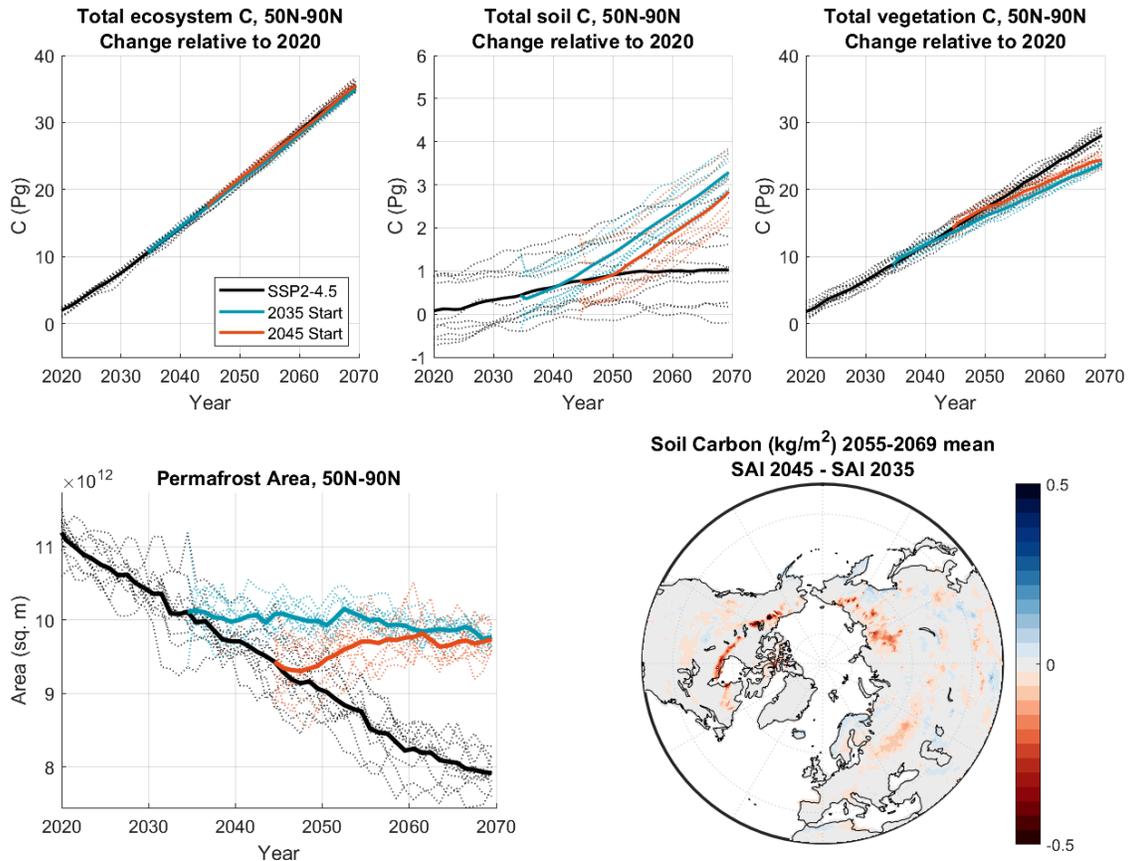

*Figure 7: Time series of total land Carbon (top left), soil Carbon (top middle), vegetation Carbon (top right), and permafrost area (bottom left) north of 50°N for the 3 scenarios. Permafrost is defined as an area where the active layer thickness is less than 3m year-round. Bottom right shows the spatial difference in soil carbon between SAI 2045 and SAI 2035, averaged from 2055-2069.*

Figure 8 shows the potential temperature in parts of the ocean in contact with the Antarctic ice sheet, which are thus important in controlling ice sheet melting. The ocean temperature in the entire shelf ocean increases steadily throughout the simulation duration in SSP2-4.5. In the East Antarctic Ice Sheet (EAIS) region, SAI halts this temperature increase within a few years of deployment, but in the Amundsen Sea/Bellingshausen Sea (ASBS) region, SAI slows but does not stop this increase by 2070. This arises because of the SAI-induced changes in atmospheric circulation and surface wind shear in the Antarctic region and its subsequent impacts on ocean circulation and upwelling, as detailed in Goddard et al. 2023. In the EAIS region, SAI 2045 recovers the ocean temperature to the same level as SAI 2035 by 2070. However, this will not recover the parts of the Antarctic ice sheet that are lost as a result of heightened temperatures between 2035 and 2070. In the ASBS region, the 10-year delay in the SAI deployment has only a small effect on the shelf ocean temperature. A much larger intervention is likely required to preserve the West Antarctic Ice Sheet (Goddard et al. 2023).

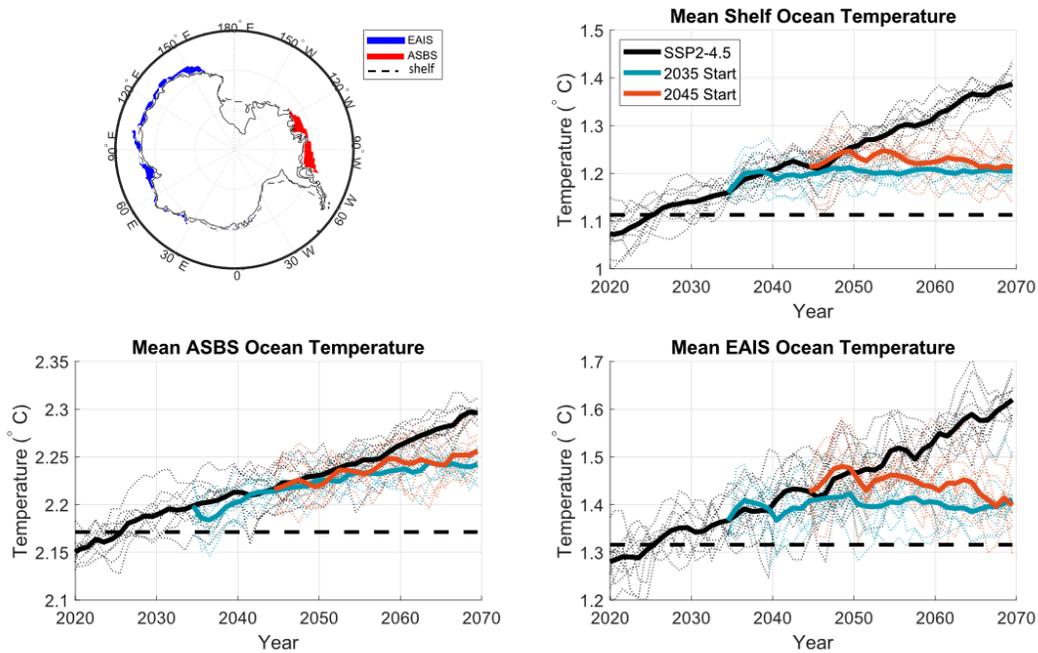

*Figure 8: Top left defines the regions referenced by the other panels. The shelf ocean is defined as the area where the ocean depth is less than 1500m, which is outlined with a dashed black line. The areas in blue and red are subregions of the shelf ocean - the East Antarctic Ice Sheet (EAIS) and Amundsen Sea Bellingshausen Sea (ASBS), respectively. The other panels show time series of mean potential temperature from 200m-1000m depth in the 3 regions. The dashed black lines represent the values from the reference period (2015-2035).*

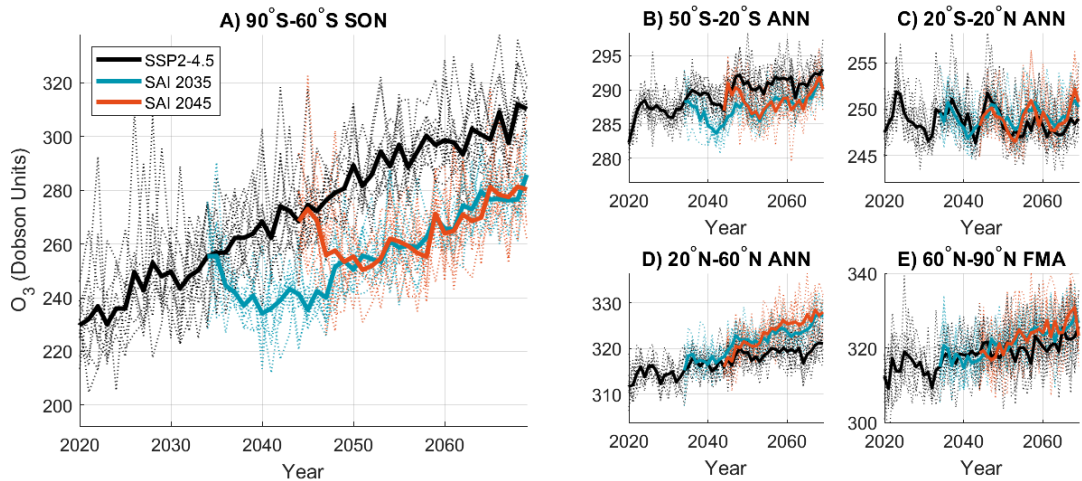

*Figure 9: Total ozone column in four different regions vs time for the four simulations. From left to right, Figure (9A) - ozone column south of 60°S, averaged from September through November. Figure (9B) - ozone column from 50°S to 20°S, annually averaged. Figure (9C) - ozone column from 20°S to 20°N, annually averaged. Figure (9D) - ozone column from 20°N to*

*60°N, annually averaged. Figure (9E) - ozone column from 60°N to 90°N, averaged from March through May.*

Figure 9 shows the time series of the total column ozone in various regions. The SAI-induced ozone response differs greatly with latitude. In the Antarctic during springtime, halogen-induced ozone depletion is the strongest due to the strong and cold polar vortex favoring polar stratospheric cloud formation and halogen activation. With the long-term reduction in stratospheric halogens, as well as the GHG-induced changes in stratospheric temperatures and transport, ozone is projected to increase under SSP2-4.5 over the next several decades, and this effect is strongest in the SH high latitudes (Figure 9A).

Under SAI, Antarctic ozone levels decrease initially once SAI is started, level off around 5-10 years after the start date, and then continue to increase at roughly the same rate as ozone in the SSP2-4.5 simulation. The initial drop is due to the formation of particularly small aerosols for the few years after the beginning of the deployment, which are particularly efficient in halogen activation (Tilmes et al., 2008).

Both SAI scenarios see a minimum Antarctic ozone level equal to that of 10 years before SAI is deployed. In SAI 2035, the minimum level of Antarctic ozone is around the 2025 level, and in SAI 2045, the minimum level of Antarctic ozone is around the 2035 level. In the long run, however, the start date does not have a noticeable effect, due to opposing impacts from lower background halogen levels but higher stratospheric sulfate loadings for SAI 2045 strategy (compared to SAI 2035). Ozone levels are set back about 20 years in both scenarios. The trend is similar for annually-averaged ozone columns in the southern hemisphere midlatitudes, partially reflecting the contribution of the Antarctic ozone following the vortex break-up in summer, but with a much smaller magnitude (Figure 9B).

In the tropics, ozone changes due to SAI are small compared to the natural variability from the 11-year solar cycle.There appears to be a small SAI-induced increase in ozone in this region after 2055, but it is small and the effect of the start date is inconclusive.

In the northern hemisphere, SAI increases ozone concentrations. This is a result of the corresponding acceleration of the deep branch of the Brewer-Dobson circulation, transporting more ozone poleward, as the results of aerosol-induced stratospheric heating (Bednarz et al. 2023b). This effect accounts for the increase in total ozone columns simulated under SAI in the NH mid-latitudes compared to SSP2-4.5. Owing to the larger injection rates needed in the delayed start case to reach the same global mean surface temperatures (Figure 2A), giving rise to larger stratospheric aerosols levels and the resulting stronger lower stratospheric heating (not shown), the acceleration of the BDC is stronger in the delayed start case, leading to larger increases in the NH total ozone columns simulated in the NH mid-latitudes.  In the Arctic, while the halogen-catalyzed ozone depletion inside the polar vortex is an important driver of the Antarctic ozone,  this effect is not as strong  in the northern hemisphere due to generally weaker, warmer and more variable NH polar vortex. As a result, in these simulations, changing the start date of SAI does not seem to play a major role in influencing the long-term evolution of Arctic springtime ozone, although the interannual variability of the simulated springtime ozone columns is high (Figure 9E).

## IV. Discussion
### A. Conclusions from this study

The 10-member ensembles of CESM2-WACCM6 simulations under various SAI scenarios show that this strategy could be successfully used to reduce global mean temperatures to offset surface warming. The various scenarios that we considered here show that in order to reach a given temperature target, SAI could be started when that target temperature is reached, and slowly ramped up to maintain it, or it could be deployed successfully later. However, the latter strategy would require a more rapid increase in SAI to achieve the target temperature. In this latter case of delayed start of intervention, we have demonstrated here that a lower net radiative forcing (-30%) and thus a higher $SO_2$ injection rate (+20%) is required. This higher injection rate is needed even after global mean surface temperature converges to the target level. This would necessitate a larger $SO_2$ supply chain and a larger fleet of aircrafts, increasing the cost of a SAI deployment. Furthermore, the additional atmospheric sulfate load would bring with it more of the potentially undesirable physical effects associated with SAI, including stratospheric heating and acid rain.

In the model used in this study, CESM2-WACCM6, the surface air temperature, precipitation, and Arctic sea ice area are largely similar for the two start dates during the 10-25 years following the delayed SAI start, although longer simulations would be needed to assess any longer-term impacts. However, there are some climate impacts that cannot be reversed fully with SAI. Additional permafrost is lost with a delayed start, and although it can be refrozen, the carbon released into the atmosphere does not get automatically recaptured, although this may be offset by increases in ecosystem carbon. Likewise, although the Antarctic shelf ocean can be recovered to the same temperature under both scenarios, any portion of the ice sheet lost during the overshoot would not be restored. With a relatively short delay of ten years, the Atlantic Meridional Overturning Circulation in this model is recovered back to the same state even with the delay, but that is not necessarily what would happen with a longer delay (e.g. Pflüger et al) or in the real world. Ultimately, these risks will need to be weighed against reasons for delaying a decision to deploy, such as the increased understanding that would be gained with another decade of research.

Our study demonstrates that there is benefit to delaying SAI deployment in terms of the ozone layer. The delayed deployment of SAI allows the ozone layer to recover further. The deployment strategy used in this study causes Antarctic springtime ozone levels to dip to levels from 10 years before the start of deployment before continuing to recover, so delaying SAI means the minimum ozone levels during deployment wouldn't be as low. However, the start year of deployment does not seem to influence long-term ozone levels in the region, although the results could be sensitive to the SAI strategy chosen and the model used.

### B. Limitations and future work

There is much more research to be done on how different SAI scenarios have different outcomes. Firstly, this study is a comparison between just two scenarios with a 10-year difference in start date. Changing this delay may yield different conclusions; for example the 70-year delay in Pflüger et al (2024) shows marked differences in AMOC state and an inability to restart deep convection in the North Atlantic. Even with the same delay, changing how fast the surface climate is cooled back down to the target temperature could yield different conclusions. It is also important to evaluate a wider range of scenarios with differences aside from just the start date. Temperature targets, background emissions, and deployment inconsistencies are all important factors, and knowing how all of them affect climate outcomes is important for policy makers.

Additionally, the scenarios in this study use the same deployment strategy. The same ten year delay with a different deployment strategy may result in different conclusions.

Even with just the simulations used in this study, there is far more analysis that could be done. For example, only three potential irreversibilities - Arctic sea ice, Arctic permafrost, and the Antarctic ice sheet - were analyzed in this study. There are far more potential irreversibilities, and output from these simulations could be used to shed light on how delaying SAI deployment could affect their associated risks. Additionally, this study only analyzed a handful of output variables in CESM2. There are far more variables, and their analysis may yield interesting conclusions that this study failed to identify.

Another important limitation is that this study uses a single climate model, CESM2 with WACCM6 as its atmospheric component. It would be valuable to conduct similar studies in other Earth system models to evaluate the robustness of conclusions.

**Acknowledgements**

The authors would like to acknowledge high-performance computing support from Cheyenne (https://doi.org/10.5065/D6RX99HX) provided by NSF NCAR's Computational and Information Systems Laboratory, sponsored by the National Science Foundation, and from Amazon Web Services (AWS). Support for E. Brody, D. G. MacMartin and D. Visioni was provided in part by the Cornell Atkinson Center for a Sustainable Future and in part through Silver Lining's Safe Climate Research Initiative. Support for E. M. Bednarz was provided by the National Oceanic and Atmospheric Administration (NOAA) cooperative agreement NA22OAR4320151 and the NOAA Earth's Radiative Budget (ERB) initiative. Support for B. Kravitz was provided in part by the National Science Foundation through agreement SES-1754740, NOAA's Climate Program Office, Earth's Radiation Budget (ERB) (Grant NA22OAR4310479), and the Indiana University Environmental Resilience Institute. The Pacific Northwest National Laboratory is operated for the U.S. Department of Energy by Battelle Memorial Institute under contract DEAC05-76RL01830. The CESM project is supported primarily by the National Science Foundation. This work was in part supported by the NSF National Center for Atmospheric Research (NCAR), which is a major facility sponsored by the U.S. National Science Foundation (NSF) under Cooperative Agreement 1852977 and by SilverLining through the Safe Climate Research Initiative. The data used in this study is from the ARISE simulations, which can be found at https://registry.opendata.aws/ncar-cesm2-arise/. Data from the 2045 start simulations and 2035 start 1.0 target simulations are not available yet, but will be in the foreseeable future.

# Supplemental Material

## I. Injection Nonlinearity

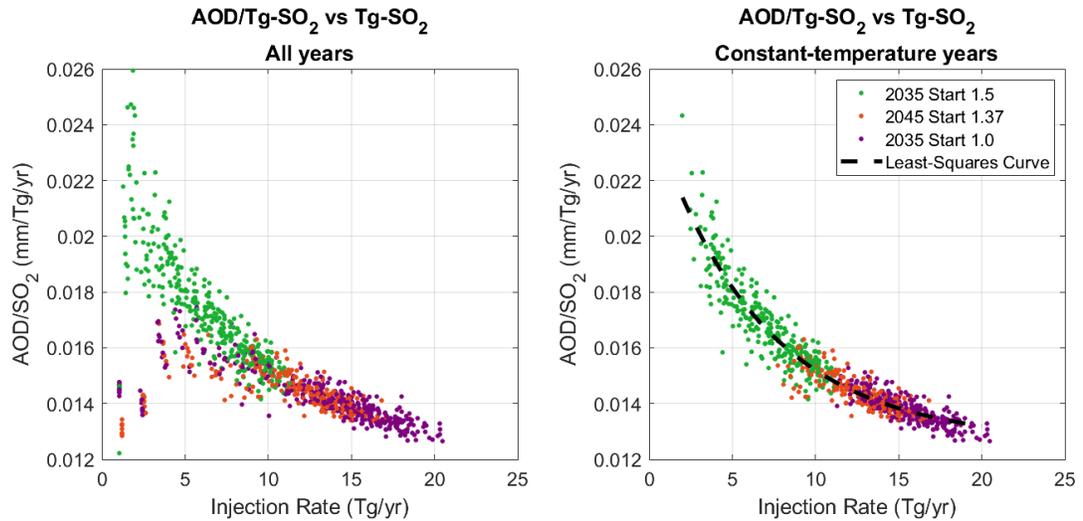

*Figure S1: AOD per Tg-SO$_2$ vs Tg-SO$_2$. Points represent individual years in individual ensemble members. Left panel shows each year in each ensemble member. Right panel shows only the years after the global mean surface temperature has converged in each simulation. This is 2040 onwards for the 2035 start 1.5° target, 2050 onwards for the 2045 start 1.37° target, and 2045 onwards for the 2035 start 1.0° target. The equation of the least-squares curve is y=0.0127+0.0119*exp(-0.1539x).*

    The relationship between SO$_2$ injection and AOD is nonlinear, as shown in figure S1. It also has time-dynamics as SO$_2$ reacts into aerosols, which then coagulate and eventually fall to the surface. This makes constructing a comprehensive emulator difficult. To deal with this, an emulator is designed to only be accurate when SAI is keeping global mean temperatures constant using the controller strategy in this study under SSP2-4.5 emissions. The data for these specific years are shown in the right panel in figure S1. The data fit a decaying exponential curve well.

    The relationship between cooling and AOD is a linear dynamic system, as shown in figure S2. All three SAI simulations have overlapping results despite having different levels of cooling, demonstrating linearity. The response follows that of a first-order system well in the 35-year time frame that was simulated.

    From these two relationships, a model could be created relating desired cooling to needed injection.

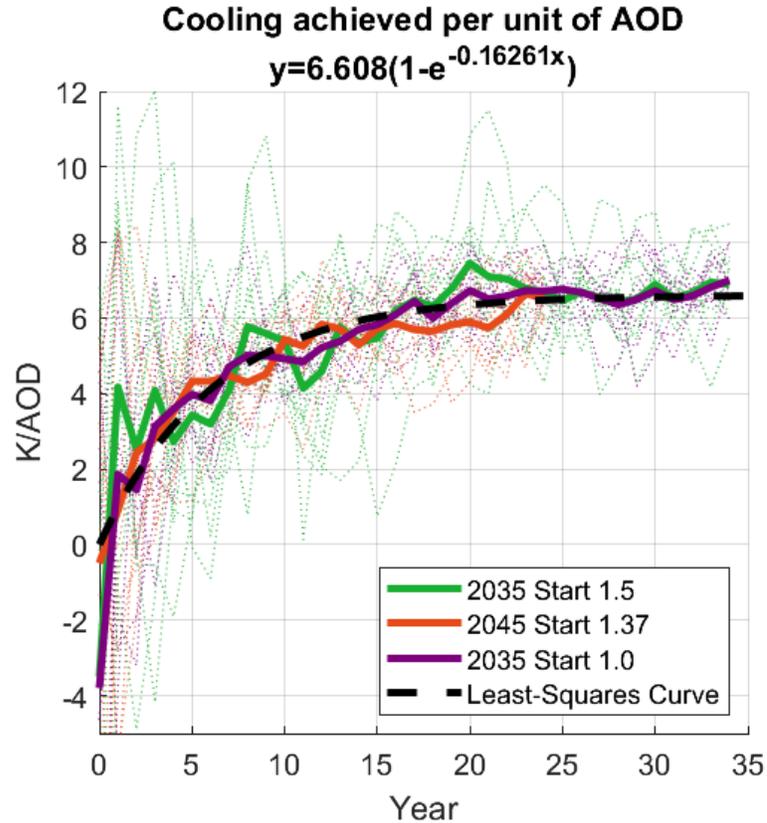

*Figure S2: Cooling achieved per unit of AOD as a function of time, with an overlayed best-fit decaying exponential. Thick solid lines are ensemble averages and thin dotted lines are individual ensemble members.*

The model used is as follows. Let $t$ denote time in years, $u(t)$ the mass of sulfur dioxide injected in year $t$, $a(t)$ the AOD in year $t$, and $x(t)$ represent the cooling in degrees Celsius in year $t$. Furthermore, let the function $F(t)$ represent the cooling achieved by a unit of AOD in year $t$, and $G(u)$ represent the AOD per Tg-$SO_2$ achieved by injection $u(t)$. From this, the cooling in a given year can be represented as follows.

$$x(t) = F(t) * G(u(t)) * u(t)$$

This can be rearranged to solve for need injection $u(t)$.

$$u(t) = x(t) * [F(t) * G(u(t))]^{-1}$$

The values of $G(u)$ and $F(t)$ are shown in figures S1 and S2, respectively. They are as follows.

$$G(u) = 0.127 + 0.0119e^{-0.1539u}$$

$$F(t) = 6.608(1 - e^{-0.16261t})$$

Given these $G(u)$ and $F(t)$, the needed injection $u(t)$ for a given cooling $x$ and year $t$ can be solved iteratively. The results are shown with a dotted line in figure S3 with 5-year smoothing. The smoothing is necessary as the desired cooling has large interannual variability. The results are very similar to performing a linear interpolation of the two SAI simulations with a 2035 start date.

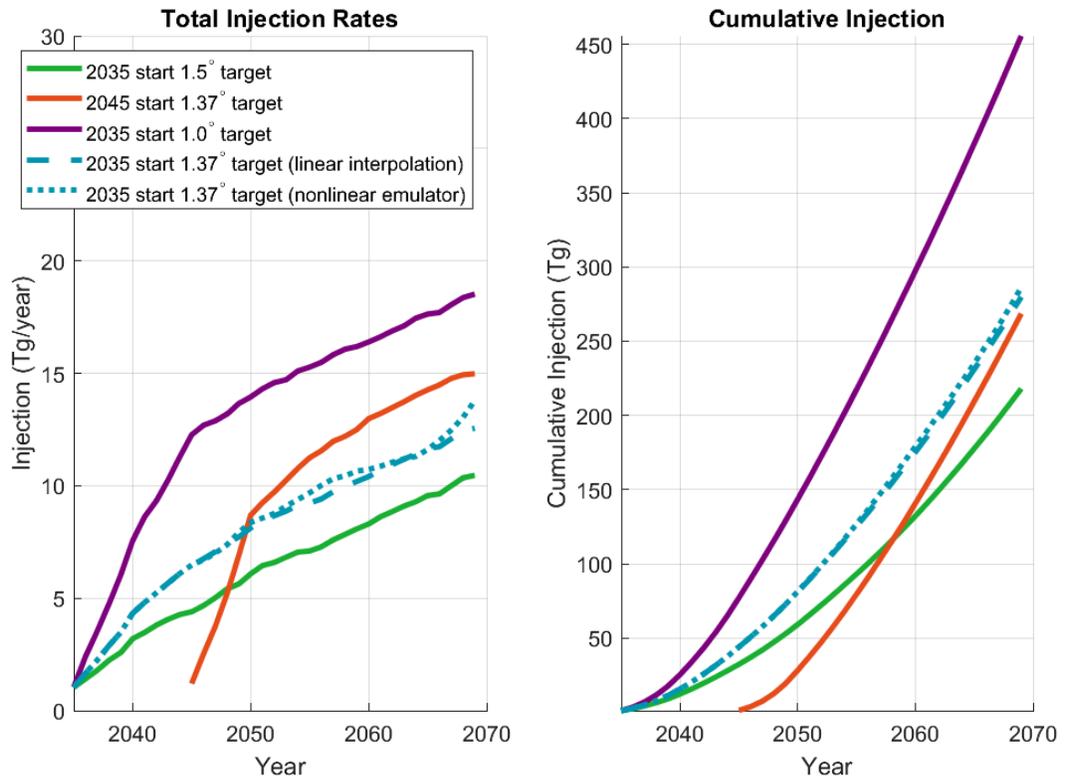

Figure S3: Total injection rates and cumulative injection for the 3 CESM SAI scenarios and the constructed SAI 2035 scenario. SAI 2035 injections are estimated with linear interpolation (dashed line) and a nonlinear emulator (dotted line). The nonlinear emulator is shown with 5-year smoothing.

**II.  Additional Figures**

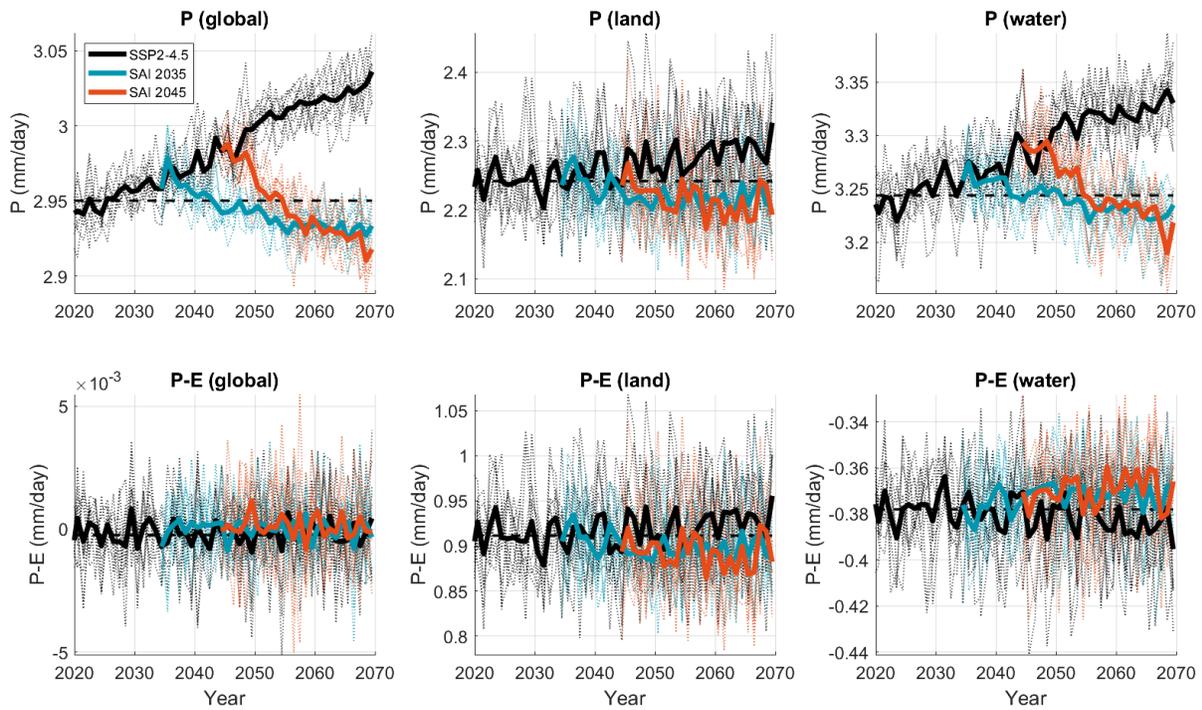

*Figure S4: Precipitation (top) and precipitation minus evaporation (bottom) globally (left), over land (middle) and over ocean (right). Dashed black lines represent values from the reference period.*

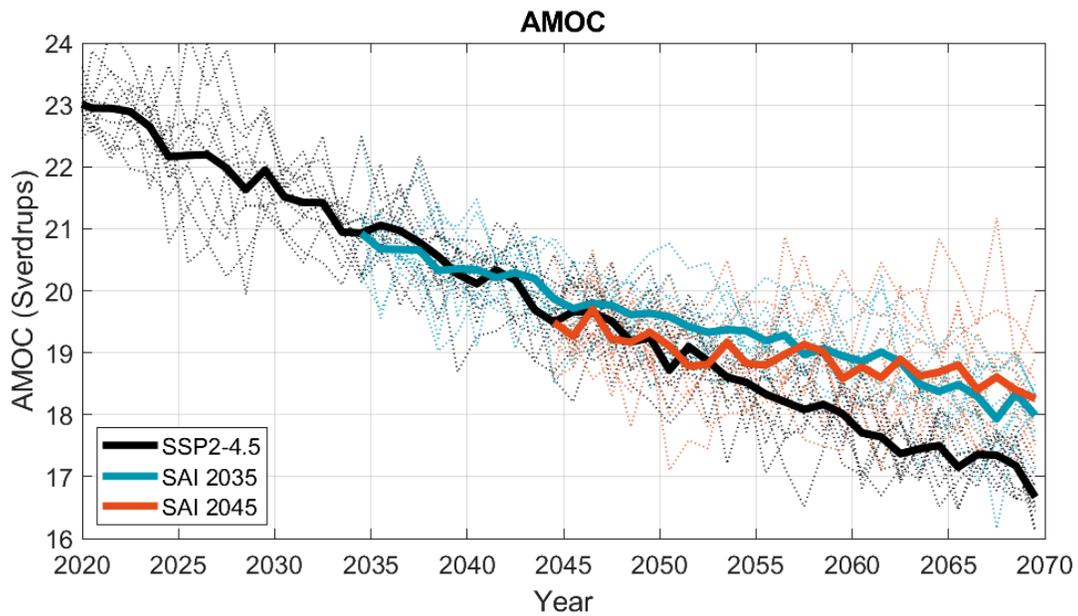

*Figure S5: Atlantic Meridional Overturning Circulation*

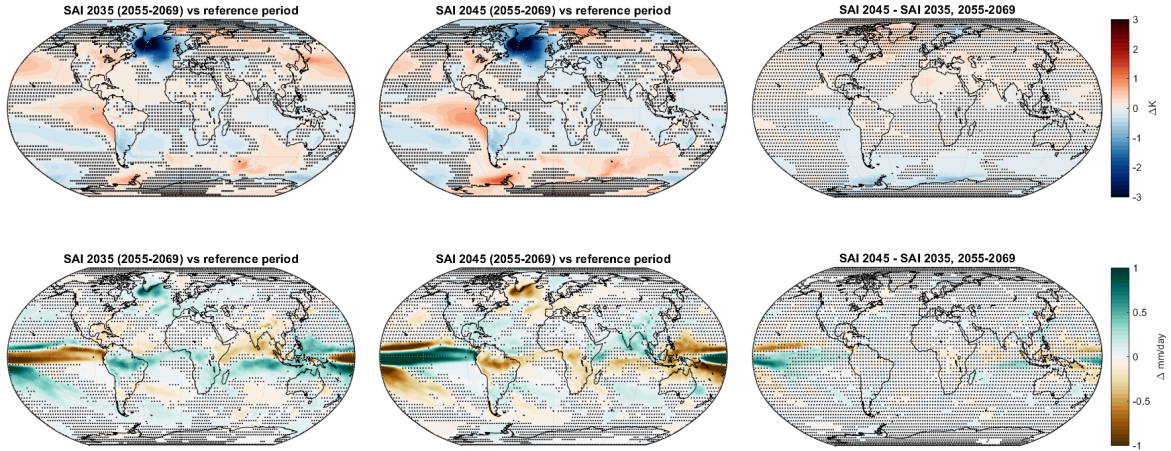

*Figure S6: Average temperature difference between: SAI 2035 2055-2069 and SSP2-4.5 2015-2035 (top left),SAI 2045 2055-2069 and SSP2-4.5 2015-2035 (top middle), and SAI 2045 2055-2069 and SAI 2035 2055-2069 (top right). Bottom row is the same but for precipitation. Statistically insignificant areas are stippled.*

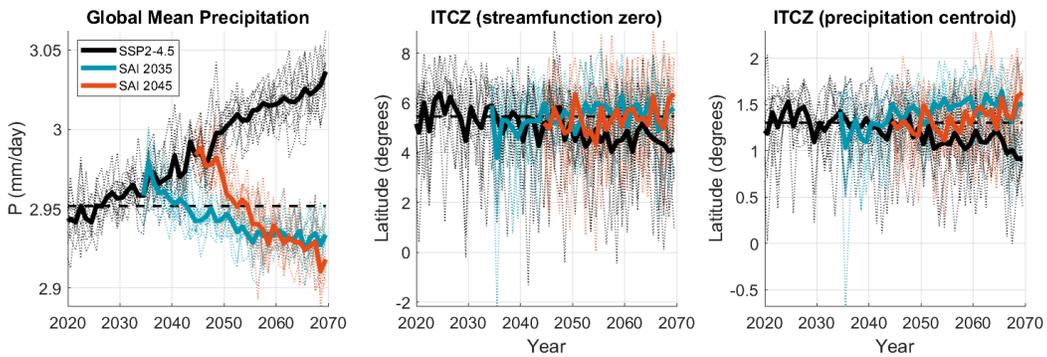

*Figure S7: Global mean precipitation (left) ITCZ as the meridional streamfunction zero (Haigh et al. 2005) (center) and precipitation centroid from 20°S to 20°N (Frierson and Hwang 2012) (right) vs time for the three scenarios. Black dashed lines are values from the reference period (2015-2035)*

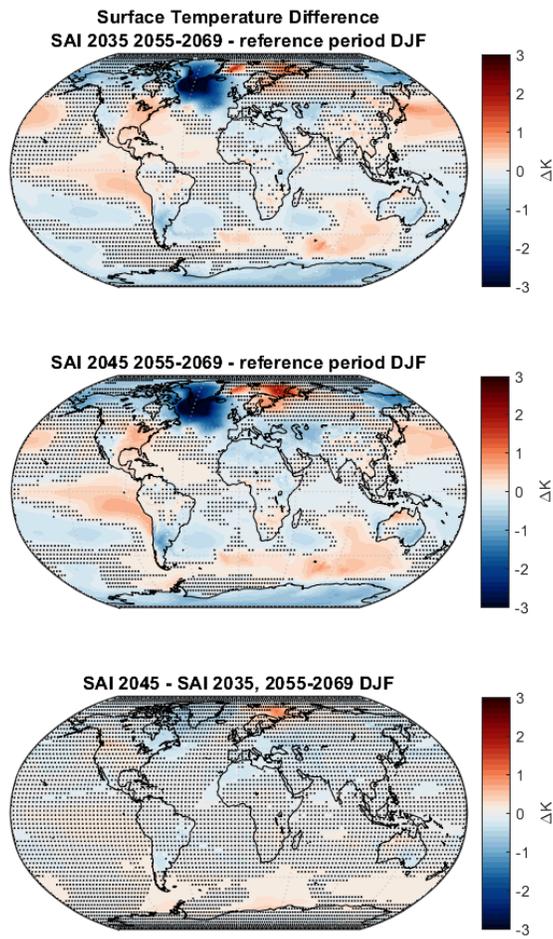

*Figure S8: December through February (DJF) mean temperatures differences between: SAI 2035 2055-2069 and SSP2-4.5 2015-2035 (top),SAI 2045 2055-2069 and SSP2-4.5 2015-2035 (middle), and SAI 2045 2055-2069 and SAI 2035 2055-2069 (bottom).*